\documentclass[10pt, conference, compsocconf]{IEEEtran}
\usepackage{cite}

\ifCLASSINFOpdf
   \usepackage[pdftex]{graphicx}
   \graphicspath{{Figures/}} 
   \DeclareGraphicsExtensions{.pdf,.jpg,.png,.jpeg}
\else
\fi
\usepackage[cmex10]{amsmath}
\usepackage[caption=false,font=footnotesize]{subfig}
\begin{document}
%
\title{Research on Damage Analysis of Key Parts of UAV Flight Control System}



\author{\IEEEauthorblockN{1st Tianshun Li}
	\IEEEauthorblockA{Automation College\\
		Northwestern Polytechnical University\\
		Xian, China\\ 
		Email: AutomationLTS@mail.nwpu.edu.cn} \\
	
	\IEEEauthorblockN{3rd Ben Xiao}
	\IEEEauthorblockA{UAV R \& D\\
		Lyncon Inc.\\
		Xian, China\\
		Email: sales@lyncon.cn}\\
	
	\IEEEauthorblockN{5th Shiyu Hao}
	\IEEEauthorblockA{Automation College\\
		Northwestern Polytechnical University\\
		Xian, China\\
		Email: AutomationLTS@mail.nwpu.edu.cn}\\
	
	\IEEEauthorblockN{7th Xuetong Wang}
	\IEEEauthorblockA{Automation College\\
		Northwestern Polytechnical University\\
		Xian, China\\
		Email: AutomationLTS@mail.nwpu.edu.cn}
	
	\and
	\IEEEauthorblockN{2nd Huaimin Chen*}
	\IEEEauthorblockA{Science and Technology on UAV Laboratory\\
		Northwestern Polytechnical University\\
		Xian, China\\
		Email: qingpei@nwpu.edu.cn}\\
	
	\IEEEauthorblockN{4th Hao Li}
	\IEEEauthorblockA{UAV R \& D\\
		Lyncon Inc.\\
		Xian, China\\
		Email: sales@lyncon.cn}\\
	
	\IEEEauthorblockN{6th Di Hai}
	\IEEEauthorblockA{Department of Simulation Test \\
		Lyncon Inc.\\
		Xian, China\\
		Email: sales@lyncon.cn}\\
}


%



\maketitle

\begin{abstract}
A set of hardware in the loop simulation methods based on the UAV model is proposed to create fault data, which is used to judge the parts where faults happen. Actual flight experimental data is utilized to prove the reliability of Simulink models. Then a series of typical faults with various amplitudes are injected into different channels of UAV parts in hardware in the loop simulation platform. Fault data is created this way, and the effect on UAV flight and task/control can be obtained through damage analysis. Typical fault characters are extracted, and those parts that have faults can be analyzed and judged. We can also know the trend that faults will develop and conclude the reasons for faults based on exterior performance, which supports precise attack and performance evaluation techniques.

\end{abstract}


%
\IEEEpeerreviewmaketitle

\section{Introduction}
UAV is a complex system that contains mechanics, electronics, automatic control, sensing, and computer control. UAVs are divided into two categories: fixed-wing and multi-rotor UAVs. UAVs have become a comprehensive weapon platform that offers destruction by fire, reconnaissance of information, and electronic countermeasures. Unmanned combat will be an important form of war \cite{IEEEhowto:Anti_UAV}. Fixed-wing UAVs often play a role in reconnaissance strike in the military field because they have a long voyage length, high speed, and big task load \cite{IEEEhowto:Ukraine}.

In the process of mission execution, the critical systems and components of UAV may be interfered with due to the complex and changeable battlefield environment and the continuous use of various anti-UAV equipment. That may affect the quality of mission execution and even result in a UAV crash, leading to severe consequences for strategic and tactical layouts \cite{IEEEhowto:Zhuhai Airshow}.
System simulation technology has made significant progress with the rapid development of control theory, modeling, and computer graphics technology. According to the difference in simulation models, the simulation technology can be divided into pure digital simulation, physical simulation, and semi-physical simulation \cite{IEEEhowto:Zongshu}. Semi-physical simulation refers to the real-time simulation of some real objects in the simulation experiment system's simulation loop. Part of the mathematical model is replaced with real objects, and the software and hardware of the system are required to run in real time. The advantage is that it is closer to the actual situation to obtain more accurate and reliable simulation information \cite{IEEEhowto:BUT}. Compared with pure digital simulation, semi-physical simulation is more realistic and economical than physical simulation, so it has a broader range of applications in aerospace, vehicle research and development, and other fields \cite{IEEEhowto:hardware in the loop}\cite{IEEEhowto:Semi-physical}.

In this paper, medium-sized fixed-wing UAVs are taken as the research object. Typical medium-sized fixed-wing UAVs are modeled after “Coyote” UAVs, developed after “Coyote” tube-launched UAVs of Raytheon. They have similar design concepts, system architecture, and aerodynamic shapes. As a typical tube-launched UAV with multiple platforms, the imitation “Coyote” UAV can be equipped with various loads and swarm operations. Whether it is launched by air, land, or water surface, it is a significant threat. Therefore, it is of representative significance to study its failure mode. If the actual flight is directly used to obtain the simulation test data, the research and development cycle is long, and the development cost is high. Therefore, the design and development of the semi-physical simulation platform for medium-sized fixed-wing UAVs are conducive to increasing the accuracy and reliability of the simulation.

In the early 21st century, unmanned aerial vehicles were mainly applied to replace manned piloting in harsh environments, aiming at meeting the technical combat index. There are no strict requirements on the service life of UAVs \cite{IEEEhowto:Small UAV}. With the continuous deepening of intelligence and extensive data integration, the unit cost of UAV equipment keeps rising. Zhong et al. proposed an adaptive three-level Kalman filter \cite{IEEEhowto:R2} to reduce the computational burden and establish the dynamic model of UAV, thus realizing the fault diagnosis of multi-rotor actuator in different scenarios. Heredia and Ollero A proposed an input-output model with a Romberg observer \cite{IEEEhowto:Kalman filter} to estimate the measured values of sensors and actuators, thus realizing the detection of UAV sensor faults and verifying them with real flight data. The damage effect analysis method of key components proposed in this paper is conducive to timely confirming whether the UAV has a fault and the fault location according to the external performance of the UAV. The fault characteristics are summarized to achieve the purpose of extending the service life of UAVs.

The fault data of drones is of great significance for fault detection \cite{IEEEhowto:Data Generation} . Due to the lack of real fault data, the previous research has proposed three types of methods for generating fault data for unmanned aerial vehicles: full digital simulation, semi-physical simulation, and physical simulation. Physical simulation is the earliest method of generating fault data, but its drawbacks include long simulation time and difficulties in parameter modification. With the development of computer technology, all digital simulation has gradually replaced physical simulation, with the advantages of low cost and easy implementation. Representative digital simulation software includes Flightgear and MATLAB/Simulink. Flightgear is an open source flight simulator released on the internet. Flightgear provides a professional flight dynamics model that can set up various scenarios, weather, and aircraft types in simulation \cite{IEEEhowto:Simulation System Based on Simulink}, it has strong flight visualization and interactive simulation capabilities, and is widely used in simulation modeling of unmanned aerial vehicles and design research of visual simulation systems \cite{IEEEhowto:HILS}\cite{IEEEhowto:visualizing simulation}. L Wang and others built a joint system based on MATLAB and Flightgear to recover and generate fault data after a drone crash \cite{IEEEhowto:Wang Lu}. However, the Flightgear platform has weak processing ability for a large amount of experimental data generated during the simulation process, and lacks a systematic data management and analysis tool \cite{IEEEhowto:FlightGear}. At the same time, the selection of fault injection nodes, types, and sizes is not flexible enough, which is not conducive to fault simulation and data generation. MATLAB/Simulink has powerful data processing capabilities, and Simulink provides a kind of graphical programming language \cite{IEEEhowto:Jia Qiuling}, which is easy to operate and avoids writing and debugging a large amount of flight control code. Tousi et al. established a linear model of UAV system \cite{IEEEhowto:R1} and proposed a fault detection and location algorithm based on robust observer. Natural factors such as strong wind, heavy snow and frost are considered to improve the accuracy of fault detection. The effectiveness of our proposed approach is illustrated and demonstrated through digital simulation results. However, full digital simulation’s credibility is insufficient, and the established model does not match the actual data very well. Semi-physical simulation integrates a portion of the actual hardware of the aircraft into the simulation system, replacing the physical model with a portion of the physical model, avoiding modeling difficulties and making the simulation system closer to the actual situation, improving the reliability of fault data. Semi-physical simulation combines the advantages of high reliability in physical simulation and low cost in digital simulation, avoiding the disadvantages of high cost in physical simulation and low reliability in digital simulation, it can flexibly simulate various fault modes \cite{IEEEhowto:Design}. DSPACE and xPC Target have been widely used in laboratory research \cite{IEEEhowto:dSPACE} \cite{IEEEhowto:xPC}.
Considering about the pros and cons, this article will use a semi-physical simulation method to generate fault data for unmanned aerial vehicle flight control systems. Compared to the previous work, the fault generation method proposed in this article can simulate the randomness of the occurrence time of fault, fault location, and fault mode, and the switching between fault modes and fault locations is simple. It can obtain fault data from different time periods, different parts, and different fault modes, solving the problem of lacking fault samples for unmanned aerial vehicles.


\section{UAV Modeling}
The simulation model of the flight control system is the basis of fault simulation. This chapter introduces the mathematical model of the UAV. It verifies the matching between the UAV model and the actual flight data to enhance the credibility of the semi-physical simulation.

\subsection{Drone Model}
Based on the aerodynamic data of the true UAV, combined with Newton's rigid body dynamics and kinematics theory, the 6-DOF nonlinear model and dynamic system model of UAV were established.

\begin{figure}[!t] 
\centering
\includegraphics[width=3.25in]{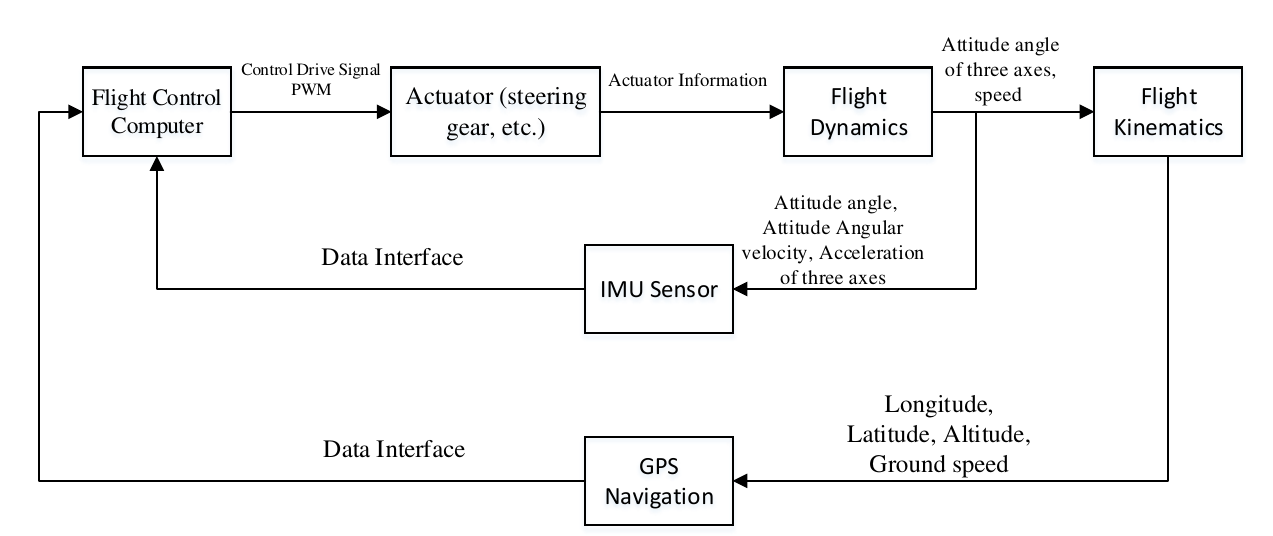}
\caption{Schematic diagram of the control system of the UAV.}
\label{fig_NO.1}
\end{figure}

The American “Coyote” UAV is taken as the research object in this paper. The physical parameters of the mathematical model established are consistent with those of the true UAV, and the control strategy is also wholly consistent. The motion mode of UAV mainly includes three linear and three angular motions, that is, spatial position change and attitude change. Generally, the 6-DOF motion state of a UAV is described by selecting the appropriate coordinate system and corresponding motion parameters. Figure \ref{fig_NO.1} shows the control system of the UAV. Figure \ref{fig_NO.2} demonstrates Simulink model of the UAV.

According to the actual situation of the UAV, the following assumptions should be made before building the nonlinear mathematical model of the UAV \cite{IEEEhowto:buaa} :
\subsubsection{{}}
A UAV is a rigid body with constant mass.

\subsubsection{{}}
The ground coordinate system is considered an inertial coordinate system.

\subsubsection{{}}
Ignoring the curvature of the Earth, the “Flat Earth Hypothesis” is adopted.

\subsubsection{{}}
The gravitational acceleration is constant.

\subsubsection{{}}
In the body coordinate system, the plane $o_{b}x_{b}z_{b}$ is the plane of UAV symmetry, the UAV geometric shape is symmetrical, the mass distribution is uniform, and the inertia product $I_{xy}=I_{yz}=I_{xz}=0$.

%
%

\begin{figure*}[ht] 
	\centering
	\includegraphics[width=6.25in]{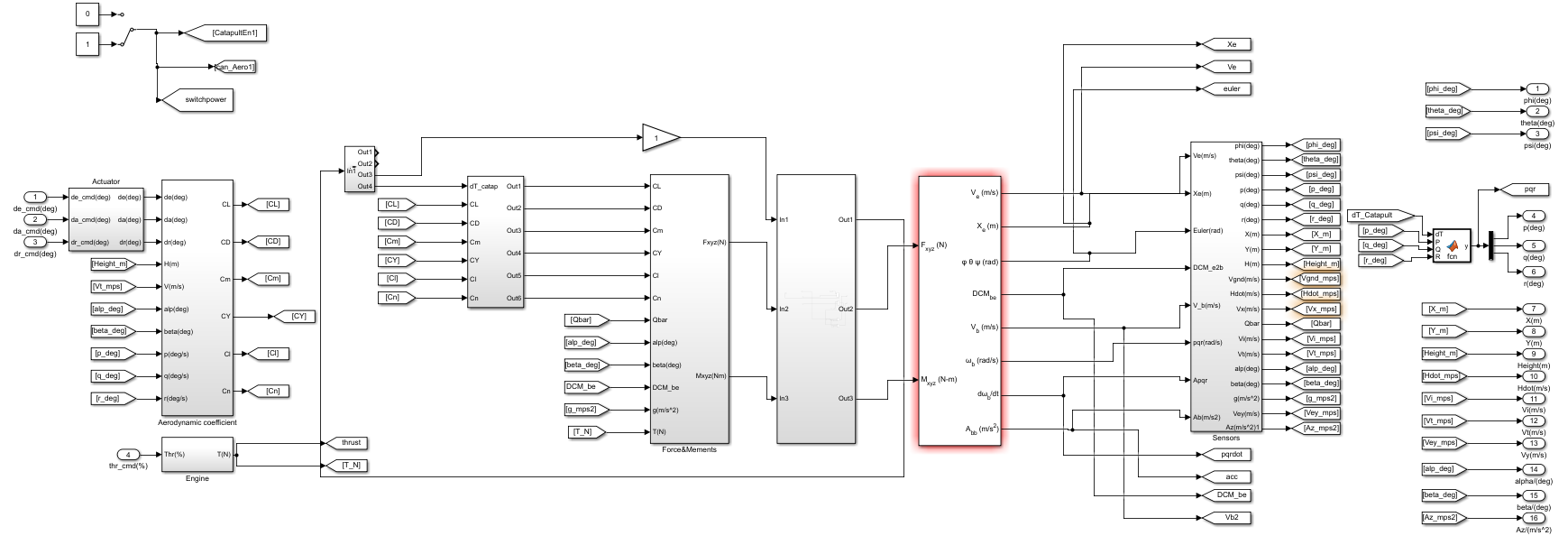}
	\caption{Simulink simulation model of UAV.}
	\label{fig_NO.2}
\end{figure*}

The Simulink simulation model of UAV in figure \ref{fig_NO.2} includes a dynamics model, propulsion system model, and aerodynamics model. The UAV dynamics model receives information such as mass, the moment of inertia, local gravity acceleration, thrust, aerodynamic force, and moment. According to the differential equations, the position, velocity, attitude angle, angular velocity, and other information of the aircraft are calculated in real-time. The aerodynamics module inputs the UAV speed, attitude, rudder deflection angle, atmospheric density, local sound speed, and other information, and interpolates the aerodynamic force and moment according to the test data and calculation data; The propulsion system module calculates thrust by interpolation according to real-time flight status and thrust test data table.


\subsection{Validation of Model}
After the establishment of the system model, it is necessary to verify the model to ensure the accuracy of the simulation model and the certainty of the simulation results and ensure that the input of the system can obtain the correct response, which is the key to flight simulation research \cite{IEEEhowto:Large Civil}.

\subsubsection{Model Matching Verification}
\begin{figure}[!t] 
	\centering
	\includegraphics[width=3.25in]{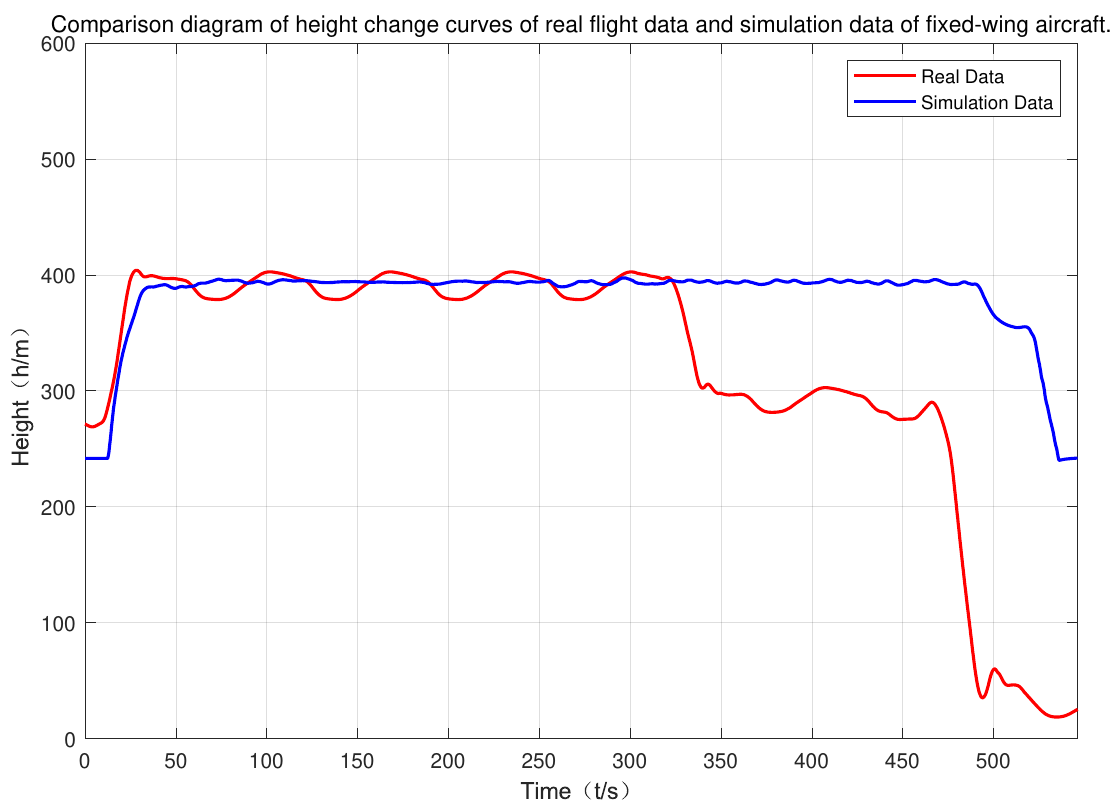}
	\caption{Comparison of height variation curves between real flight data and simulation data of fixed-wing aircraft.}
	\label{fig_NO.3}
\end{figure}

\begin{figure}[!t] 
	\centering
	\includegraphics[width=3.25in]{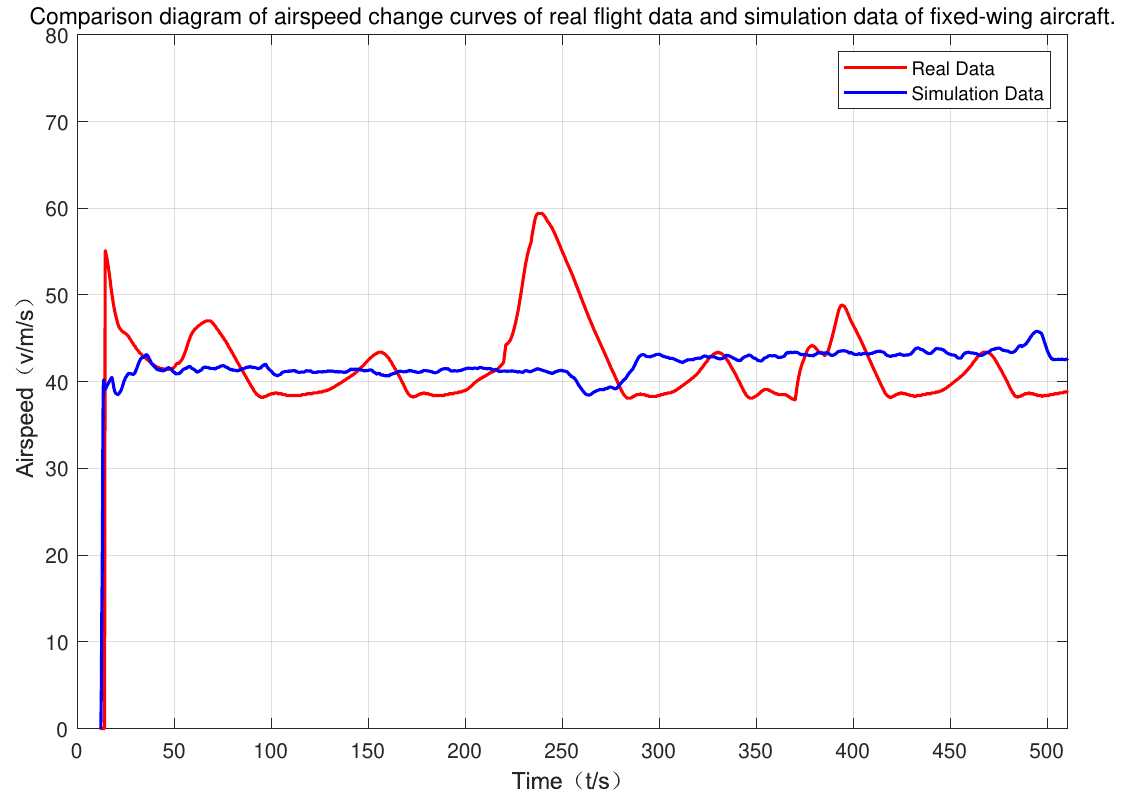}
	\caption{Comparison of airspeed curve between real flight data and simulation data of fixed-wing aircraft.}
	\label{fig_NO.4}
\end{figure}

According to the real flight data, the flight test with a cruising altitude of 150m and a cruising speed of 45m/s was carried out. The real flight altitude and speed comparison with the simulation test flight altitude and speed is shown in the figure below. According to the curve comparison, it can be seen in figure \ref{fig_NO.3} and figure \ref{fig_NO.4} that the simulation can cruise at a speed of 45m/s at the height of 150m.

It can be seen from the simulation results that the variation trend of the simulation output is consistent with the actual flight data, and the simulation model has a good match with the actual flight data. Considering that the real UAV will be affected by various noises and disturbances in the actual flight, it can be considered that the simulation model has a high matching degree with the actual UAV, which proves the correctness and practicability of the model.

\subsubsection{Dynamic Characteristic Verification}
\begin{figure}[!t] 
	\centering
	\includegraphics[width=3.25in]{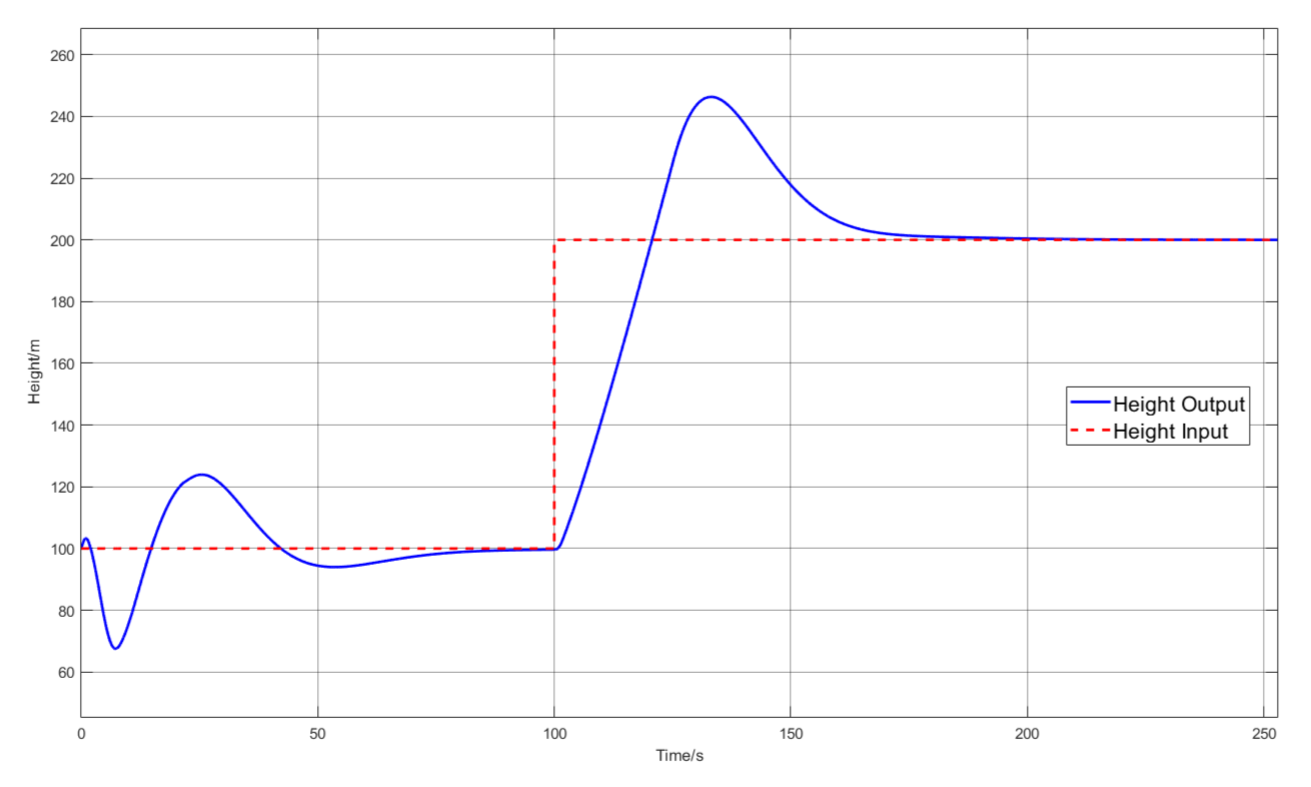}
	\caption{Step response to a longitudinal motion of the simulation model.}
	\label{fig_NO.5}
\end{figure}

Taking the longitudinal motion of the UAV as an example, the simulation time was set as 250s, and the sampling time was set as 0.001s, where a step signal with an amplitude of it was applied to the height. The step response curve of the system was obtained. As shown in the figure \ref{fig_NO.5}, the dotted line is the given step input, and the solid line is the simulation result of the height.

According to the step response curve in Figure \ref{fig_NO.5}, the maximum overshoot of the UAV system is 23.15\%, the adjustment time is 55.382s, and the steady-state error is 0. The high steady-state error tracking control is realized, which has good dynamic characteristics and proves the effectiveness of the model control law in the longitudinal control of the UAV.

\section{Composition and Fault Modeling of Flight Control System}
\begin{figure}[!t] 
	\centering
	\includegraphics[width=2.0in]{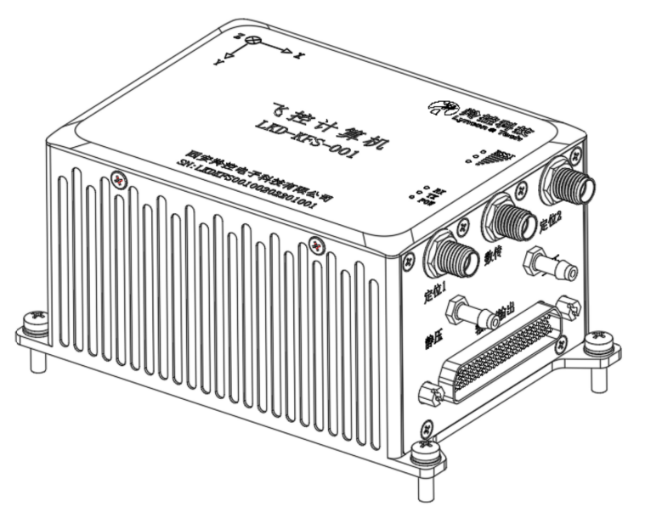}
	\caption{LKD-KFS-001 aircraft management computer.}
	\label{fig_NO.6}
\end{figure}

A physical object is the physical system of physical simulation, which is very important to the whole system. To ensure the authenticity and accuracy of the test data, the flight control computer and attitude sensor were connected to the simulation loop as the physical system, and the semi-physical simulation was carried out for the fixed-wing UAV system. LKD-KFS-001 Aircraft Management Computer is selected as the navigation flight control and aircraft management computing platform, which adopts the stacked structure design and is composed of three parts: bottom plate, processor board, and shell. The outline diagram is shown as figure \ref{fig_NO.6}.

The LKD-KFS-001 Aircraft Management Computer is used as the semi-physical simulation flight control, management, and mission computer. It has high integration, internal integration of the CPU, industrial double redundant attitude sensor, magnetometer, static and static pressure sensor and dual redundant GNSS sensors, built-in 900Mhz data link. High reliability, with full metal shielding housing and J30J aviation connector, double redundant GNSS positioning system and doubles redundant attitude sensor automatic switch. Power-on self-test and flight self-test. Power failure protection within 50ms, dual differential positioning, high control accuracy. The technical parameters are shown in table  \ref{table_AircraftComputer}.
%
\begin{table}[!t]
\renewcommand{\arraystretch}{1.3}
\caption{Technical Parameters of Aircraft Management Computer}
\label{table_AircraftComputer}
\centering
\begin{tabular}{c c}
	\hline
       

	Technical Parameters&Parameters\\
	\hline 
	Precision of Attitude&pitch /roll 0.1°\\
	Positioning Accuracy&horizontal 0.01m,vertical 0.02m(RTK)\\
	Precision of Orientation&0.2°\\
	Precision of Velocity&0.03m/s(RTK)\\
	PWM Channels&16 Roads\\
	RS422/232 Channels&6 Roads\\
	Basic Frequency&866MHZ\\
	Operating System&VxWorks\\
	Operating Temperature&[-40℃,85 ℃ ] \\
	
	\hline
\end{tabular}
\end{table}

To carry out fault injection on the above model, the fault characteristics of the steering gear and sensor are analyzed in detail. The corresponding mathematical model is established according to the common fault modes of the two, which is the basis for the subsequent fault simulation and fault injection. The implementation process of fault injection and the problems needing attention are summarized to ensure the availability of fault data.

\subsection{Analysis and Modeling of Sensor Fault Characteristics}
The sensor in UAV can measure the parameters of the flight state in real-time, which plays an important role in UAV flight. The flight control sensor can be divided into a gyroscope, accelerometer, GPS, and so on, according to the measured physical quantities. An IMU sensor typically consists of three uniaxial accelerometers and three uniaxial gyroscopes. This paper analyzes the fault characteristics of IMU and GPS, and the measured parameters are shown in Table \ref{table_SensorPara}.
\begin{table}[!t]
	\renewcommand{\arraystretch}{1.3}
	\caption{Sensor Type and Measured Parameters}
	\label{table_SensorPara}
	\centering
	\begin{tabular}{c c}
		\hline
		
		
		Sensors&Parameters Measured\\
		\hline 
		IMU&Attitude angle, Attitude angular velocity, acceleration\\
		GPS&Longitude, Latitude, Altitude, Ground Speed\\
			
		\hline
	\end{tabular}
\end{table}

Various kinds of sensor faults may happen during the actual flight of a UAV. The analysis of the characteristics of the sensor before and after the fault is beneficial to the classification and modeling of the sensor fault, laying the foundation for the subsequent fault simulation and diagnosis.

According to the causes of sensor faults, this paper analyzes and models the characteristics of GPS spoofing, weak signal, weaker signal, signal interruption fault, IMU multiplier fault, constant deviation fault, drift fault, transient signal drift, communication disconnection fault, and finally generates fault data through simulation.

\begin{figure}[!t] 
	\centering
	\includegraphics[width=3.5in]{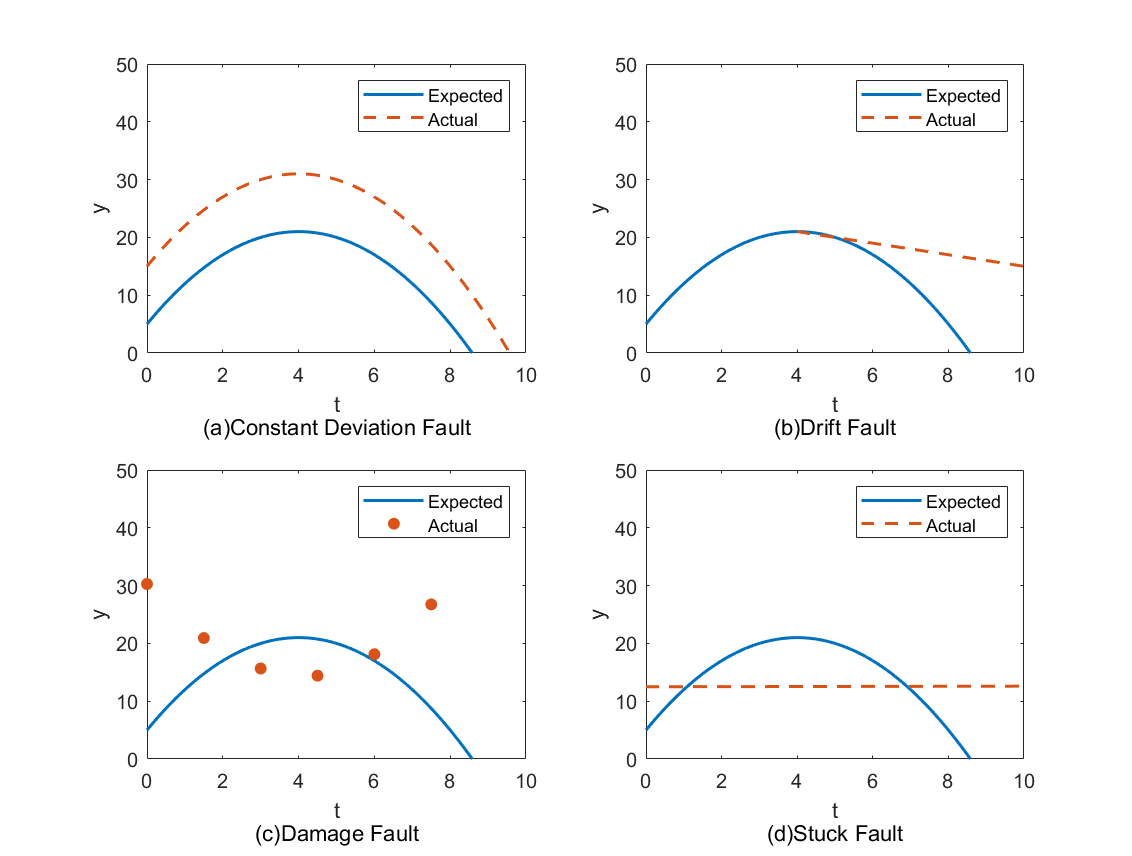}
	\caption{Fault types of sensors.}
	\label{fig_NO.7}
\end{figure}

After a comprehensive analysis of the causes and manifestations of sensor faults, the mathematical model of the sensor is expressed in the following formula \cite{IEEEhowto:Combination}\cite{IEEEhowto:data-driven}.
\begin{equation}\label{eq:sensor}
		{{y}_{s}}\left( t \right)=k\left( t \right)y\left( t \right)+d\left( t \right)  \\
\end{equation}
 Where ${{y}_{s}}\left( t \right)$ represents the given expected output value; $y\left( t \right)$ represents the actual output value; $k\left( t \right)$ represents gain; $t$represents the current moment; $d\left( t \right)$ represents the deviation generated by the output value. The table \ref{table_SensorPara} shows the importance of parameters in typical sensor fault modes. Figure \ref{fig_NO.7} shows different fault types of sensors.
 
\begin{table}[!t]
	\renewcommand{\arraystretch}{1.3}
	\caption{Sensor Fault Mode and Parameter Value}
	\label{table_gearPara}
	\centering
	\begin{tabular}{c c}
		\hline
		
		
		Failure Mode&$k\left( t \right)$ and $d\left( t \right)$ \\
		\hline 
		Fault-free&$k\left( t \right)\text{=}1$,$d\left( t \right)=0$\\
		Multiplicative Fault&$k\left( t \right)\ne 1$,$d\left( t \right)=0$\\
		Constant Deviation Fault&$k\left(t\right)=1$,$d\left(t\right)$ is constant\\
		Drift Fault&$k\left(t\right)=1$,$d\left(t\right)$ changes over time\\
		Transient Signal Drift&$k\left(t\right)=1$,$d\left(t\right)$ is a discrete value\\
		Disconnect Communication&$k\left(t\right)\text{=} d\left(t\right)=0$\\
		\hline
	\end{tabular}
\end{table}

\subsection{Analysis and modeling of steering gear fault characteristics}
\begin{figure}[!t] 
	\centering
	\includegraphics[width=3.5in]{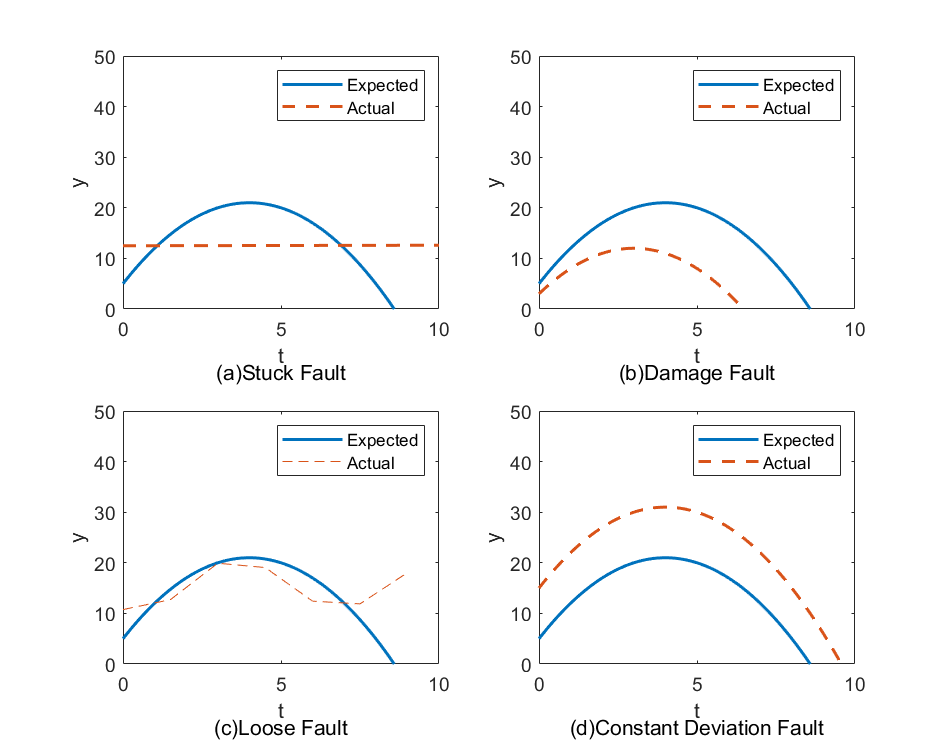}
	\caption{Fault types of the steering gear.}
	\label{fig_NO.8}
\end{figure}

This paper studies the performance forms of the fault and analyzes the characteristics of the steering gear's constant deviation fault, stuck fault, loose fault, and damage fault. The corresponding mathematical model is established. According to the established model, the fault injection and simulation are carried on, and the fault data is generated. The figure \ref{fig_NO.8} shows the fault types of the steering gear.

The mathematical model of the steering gear is shown in equation \ref{eq:gear}.
\begin{equation}\label{eq:gear}
	{{u}_{m}}\left( t \right)=k\left( t \right){{u}_{c}}\left( t \right)+d\left( t \right)  \\
\end{equation}
Where ${{u}_{m}}\left( t \right)$ represents the actual output value of the steering gear; ${{u}_{c}}\left( t \right)$ represents the given expected output value; $d\left( t \right)$ represents the deviation generated by the output value; $t$ represents the current moment; $k\left( t \right)$ represents the gain. Table \ref{table_gearfault} describes the parameters in a typical steering gear fault mode.
\begin{table}[!t]
	\renewcommand{\arraystretch}{1.3}
	\caption{Steering gear fault mode and parameter value}
	\label{table_gearfault}
	\centering
	\begin{tabular}{c c}
		\hline
		
		
		Failure Mode&$k\left( t \right)$and$d\left( t \right)$\\
		\hline 
		Fault-free&$k\left( t \right)\text{=}1$,$d\left( t \right)=0$\\
		Constant Deviation Fault&$k\left( t \right)=1$,$d\left(t\right)$ is constant\\
		Stuck Out Fault&$k\left( t \right)=0$,$d\left( t \right)$ is constant\\
		Loose Fault&$k\left( t \right)=0$,$d\left( t \right)$ changes\\
		Fault of Damage&$k\left( t \right)\subset \left( 0,1 \right)$,$d\left( t \right)=0$\\
		\hline
	\end{tabular}
\end{table}
\section{Building of Platform}
\begin{figure}[!t] 
	\centering
	\includegraphics[width=3.25in]{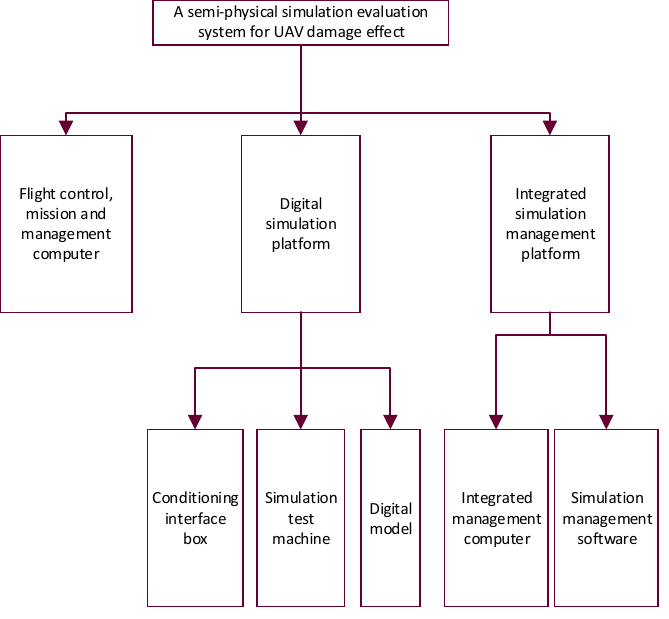}
	\caption{System composition diagram.}
	\label{fig_NO.9}
\end{figure}

The composition of the UAV damage effect semi-physical simulation evaluation system is shown in figure \ref{fig_NO.9}, which includes flight control, mission, and management computers (flight control computers), a digital simulation platform, and a comprehensive simulation management platform. Among them, the flight control computer includes the flight control computer of the UAV, which is the physical part of the semi-physical simulation system. The digital simulation platform consists of a simulation test computer, a conditioning interface module, and a digital model. The integrated simulation management platform includes the integrated management computer and its simulation management software.

The system hardware equipment is integrated into an 18U cabinet, mainly including flight control, management, and task computers, integrated management computers, simulation test computers, integrated control boxes, displays, and switches.

\begin{figure*}[!t]
	\centering
	\subfloat[Imitation Drone]
{
	\includegraphics[width=2.5in]{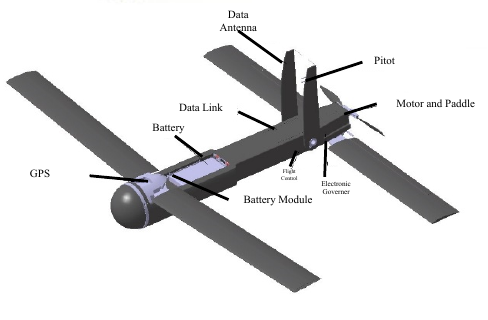}
	\label{fig_first_case}
}\hfil
	\subfloat[True Drone]
{
	\includegraphics[width=2.5in]{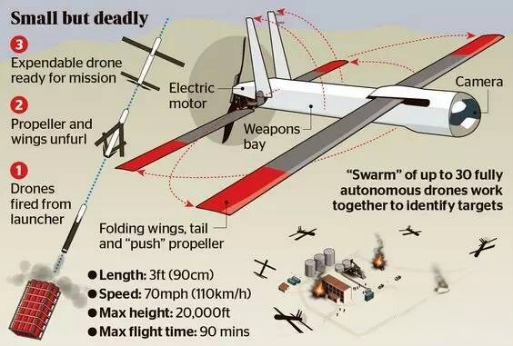}
	\label{fig_second_case}}
\caption{A comparison of a imitation and a true drone.}
\label{fig_sim}
\end{figure*}

The imitation UAV studied in this project (shown on the left in figure \ref{fig_sim}), and Raytheon Co. 's “Coyote” UAV (shown on the right in figure \ref{fig_sim}) both adopt up-down folding wings and a folded V tail, which can also be launched from the air or the ground using tubular containers.
%
%

\begin{figure}[!t] 
	\centering
	\includegraphics[width=3.25in]{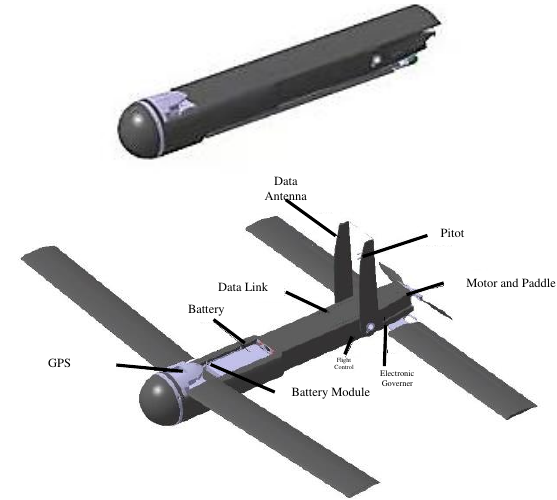}
	\caption{System composition diagram.}
	\label{fig_NO.11}
\end{figure}

The weight of the imitation UAV is about 6.5 kg (including the power device, power battery, parachute recovery system, control steering gear, and cable connector) shown in figure \ref{fig_NO.11} and table \ref{shape_para}.

\begin{table}[!t]
	\renewcommand{\arraystretch}{1.3}
	\caption{Shape parameters of the UAV}
	\label{shape_para}
	\centering
	\begin{tabular}{c c}
		\hline
		
		
		Parameters&Parameter Value\\
		\hline 
		Takeoff Weight&$9kg$\\
		Wingspan&Front 1.67$m$ Rear 1.335$m$\\
		Wing Area&Front $0.2{{m}^{2}}$
		Rear 0.135${{m}^{2}}$\\
		Body Length&1.005$m$\\
		Average Aerodynamic Chord Length&0.21$m$\\
		Cruise Speed&25$m/s$\\
		Endurance&20$\min s$\\
		
		\hline
	\end{tabular}
\end{table}

The semi-physical simulation platform is built based on the simulation test mode of the flight control computer in the loop. The flight control computer adopts the self-developed physical product, corresponding to the flight control of the imitation UAV. Digital and discrete protocol interfaces and simulation test computers manage the flight status and instructions. The flight control algorithm is solved and analyzed on a real flight control computer.

The automatic injection of the test is mainly built by the integrated management computer and the simulation testing mechanism. The integrated management computer is mainly used for simulation test management, data monitoring, tracking attitude display, data acquisition, recording, etc. The built-in resource board card of the simulation test aircraft adopts the VxWorks real-time simulation system, which is mainly used to receive the simulation instructions of the integrated management computer, perform the real-time calculation of the digital model of the UAV and the functional model of the peripherals, and send real-time flight simulation data to the integrated management computer.

\begin{figure}[!t] 
	\centering
	\includegraphics[width=3.25in]{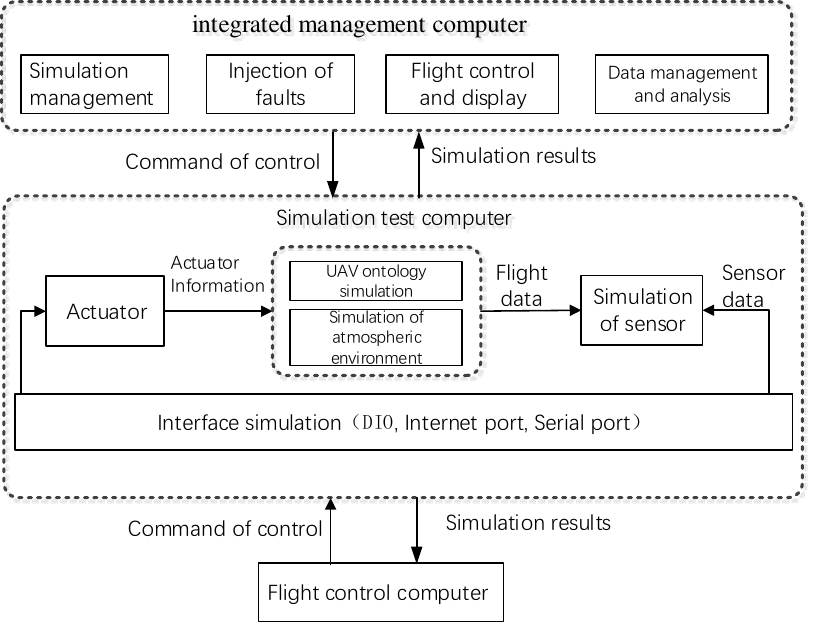}
	\caption{Schematic diagram of the test platform.}
	\label{fig_NO.12}
\end{figure}

The signal resources of the simulation test machine are connected to the flight control computer through the conditioning interface box. The integrated control box has the function of signal disconnection, conditioning, and interface adaptation shown in figure \ref{fig_NO.12}.

\section{Data generation and damage effect analysis}
\begin{figure}[!t] 
	\centering
	\includegraphics[width=3.25in]{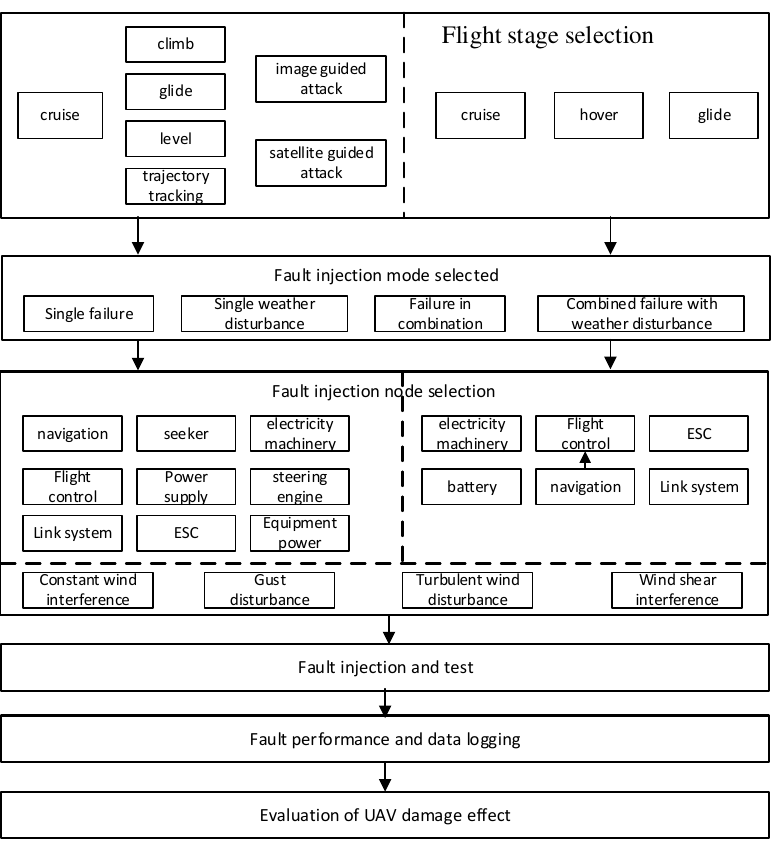}
	\caption{Test flow chart.}
	\label{fig_NO.13}
\end{figure}

This paper only analyzes the damaging effect of steering gear failure, GPS failure, and IMU failure. The test process is shown in figure \ref{fig_NO.13}, and the specific steps are as follows:

\begin{enumerate}
	\item Fault injection and test: According to the fault feature extraction strategy, the fault injection test was carried out using multiple modes and fault nodes.
	\item Test result recording and statistics: Record each test's input and output data, and make summary statistics.
	\item UAV damage effect evaluation: The damage effect evaluation of UAVs is carried out.
\end{enumerate}

\subsection{Sensor Fault}
\subsubsection{GPS Fault}

The size of the steering gear fault injected in hardware in the loop simulation is shown in table \ref{injection_GPS}.
\begin{table}[!t]
	\renewcommand{\arraystretch}{1.3}
	\caption{Specific parameters of GPS fault injection}
	\label{injection_GPS}
	\centering
	\begin{tabular}{c c}
		\hline
		
		
		GPS Fault Type&Specific Parameters\\
		\hline 
		Deception Interference&Longitude and latitude change 0.05 °,\\
		&about 5km change\\
		Weak Signal&The number of GPS satellites decreases,\\
		& from 15 to 12.\\
		Weaker Signal&The number of GPS satellites decreases,\\
		& from 15 to 7.\\
		Signal Interruption&Disconnect the GPS signal.\\
		\hline
	\end{tabular}
\end{table}

\begin{figure}[!t] 
	\centering
	\includegraphics[width=3.25in]{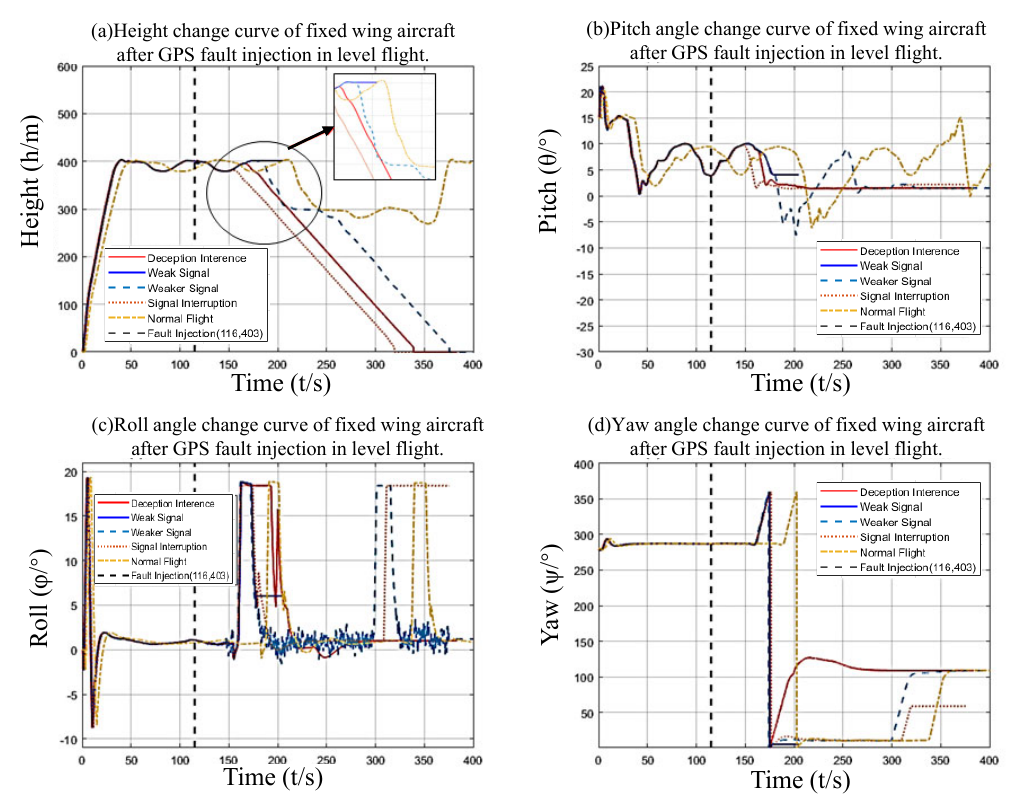}
	\caption{The influence of GPS fault on system state quantity in level flight.}
	\label{fig_NO.14}
\end{figure}

Deception jamming faults, weak signals, and signal interruption faults in level flight all affect the normal completion of the task of the UAV, but the UAV can still land and attack. The weak signal fault will not affect the flight control of the UAV.

It can be seen from figure \ref{fig_NO.14} that, after a given altitude of 400m, the UAV is subject to deception jamming for about 136 seconds in the level flight segment, GPS positioning fails, flight control is switched to IMU for positioning control, and then after about 30 seconds of IMU positioning control divergence, flight control is switched to level flight control. Since the UAV cannot be controlled at a fixed height, it gradually glides downward. After estimation, At an average speed of 40m/s, after the fault is injected in the cruise section, it can still slide forward for about 8km and attack. The distance of sliding forward depends on the flight speed and vertical speed. IMU positioning control latitude changes by 0.05 ° within one second. After a period of time, the ground station judges that GPS positioning is invalid. After receiving the fault, the flight control enters the level flight control strategy, the pitch angle is kept at about 3 °, the lateral mode remains the same as the navigation state, glides forward, and the altitude gradually reduces to the final landing for the attack. The level flight phase of GPS deception jamming influences the target of UAV's surveillance and strike mission, but it can still attack.

\begin{figure}[!t] 
	\centering
	\includegraphics[width=3.25in]{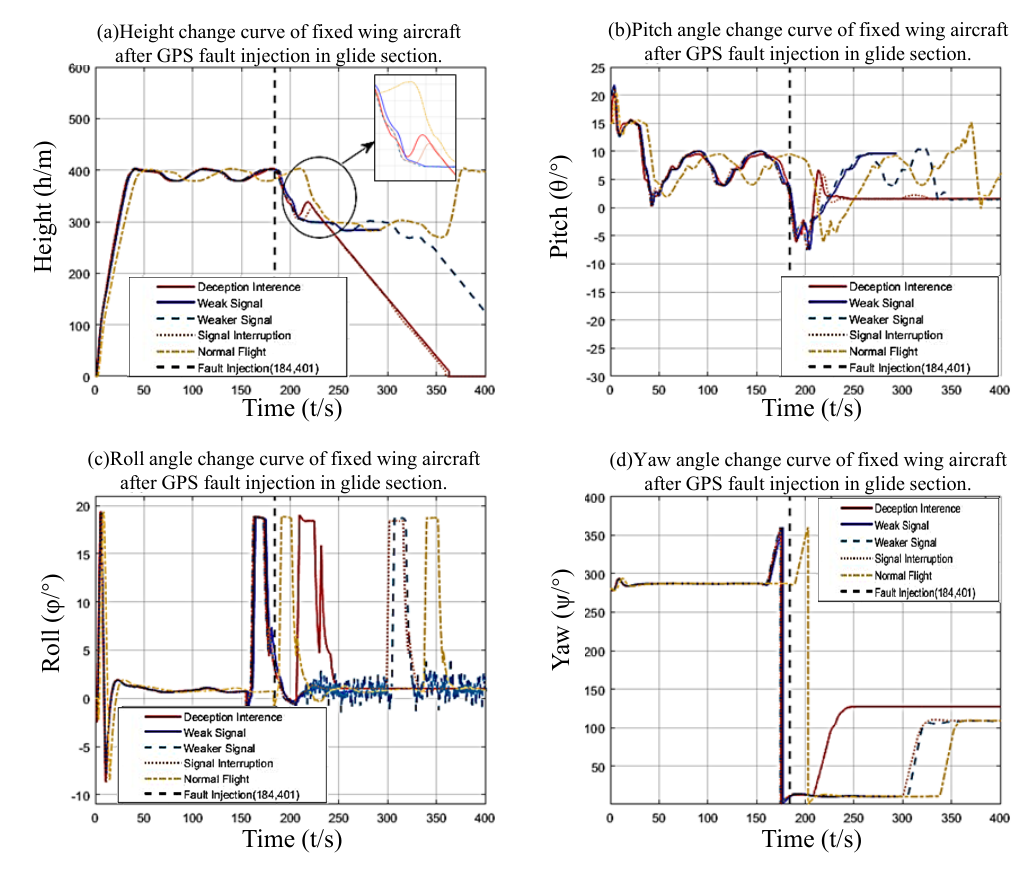}
	\caption{The influence of GPS fault on the system state quantity in glide slope.}
	\label{fig_NO.15}
\end{figure}

Deception jamming faults, weak signals, and signal interruption faults in the glide slope impact the normal completion of the task of the UAV, but the UAV can still land and attack. The weak signal fault will not affect the flight control of the UAV.

It can be seen from figure \ref{fig_NO.15} that, after a given altitude of 400m, the UAV is subject to deception jamming in the glide section for about 188 seconds, GPS positioning fails, flight control is switched to IMU for positioning control, and then after about 30 seconds of IMU positioning control divergence, flight control is switched to level flight control. Since the UAV cannot be controlled at a fixed height, it gradually glides downward. After estimation, At an average speed of 40m/s, after the fault is injected into the glide slope, it can still glide forward for about 7km and attack. The distance to glide forward depends on the flight speed and vertical speed. IMU positioning control latitude changes by 0.05 ° within one second. After a period of time, the ground station judges that GPS positioning is invalid. After receiving the fault, the flight control enters the level of the flight control strategy. The pitch angle changes from - 5 ° in the original glide state to about 3 ° in the level flight and continues to maintain. The lateral mode keeps the navigation state unchanged, glides forward, gradually decreases altitude, and finally lands to attack. GPS deception jamming climb phase influences the target of the UAV's surveillance and strike mission, but it can still attack.

\subsubsection{IMU Fault}
The IMU fault size injected in the hardware in the loop simulation is shown in table \ref{injection_IMU}.
\begin{table*}[!t]
	\renewcommand{\arraystretch}{1.3}
	\caption{Specific parameters of IMU fault injection.}
	\label{injection_IMU}
	\centering
	\begin{tabular}{c c}
		\hline
		
		
		IMU Fault Type&Specific Parameters\\
		\hline 
		Multiplicative Fault&The signal gain changes to 1.2-1.3, there is a fault on the x and z axes, and the y-axis is normal\\
		Constant Deviation Fault&The angle has a deviation between 3 ° and 5 °, \\
		&which is 3 ° on the static x-axis, 5 ° on the z-axis, and normal on the y-axis.\\
		Drift Fault&The signal drifts for a long time, about 40 seconds.\\
		Communication Disconnection Fault&The feedback signal provided by IMU is disconnected,\\
		& and the aircraft system model becomes an open-loop stochastic model.\\
		\hline
	\end{tabular}
\end{table*}
\begin{figure}[!t] 
	\centering
	\includegraphics[width=3.25in]{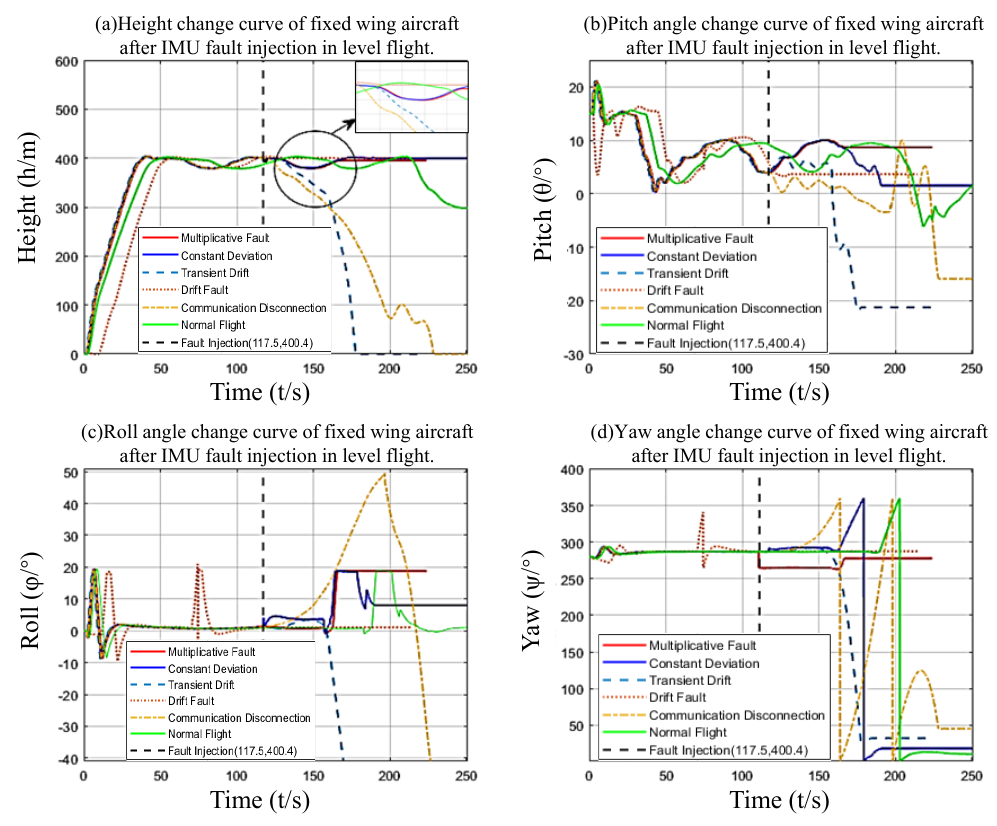}
	\caption{The influence of IMU fault on the system status in the level flight phase.}
	\label{fig_NO.16}
\end{figure}

The IMU drift fault and communication disconnection in the level flight phase impact the normal completion of the task of the UAV, but the UAV can still land and attack. Ride fault, constant deviation fault, and transient signal drift will not affect the flight control of UAVs.

The drift generated by the IMU sensor itself cannot return to the normal state for a while. The resulting error accumulation makes it difficult for the IMU to achieve long-term direction estimation, resulting in a large error between the parameters measured by the IMU and the actual.

It can be seen from figure \ref{fig_NO.16} that after a given altitude of 400m, the control strategy is to maintain the previous rudder after the UAV injects IMU drift fault in the level flight phase. Since the aircraft's attitude cannot be maintained, it can only continue to output the rudder before IMU divergence, maintain speed control, and carry out tasks. The pitch angle gradually decreases, and the aircraft lowers its head. The roll angle is also declining, and the plane turns to a left roll. The flight track changes from the original straight-line cruise to a slow glide to the left with a certain roll angle and can still attack the ground but cannot attack the target normally.

\begin{figure}[!t] 
	\centering
	\includegraphics[width=3.25in]{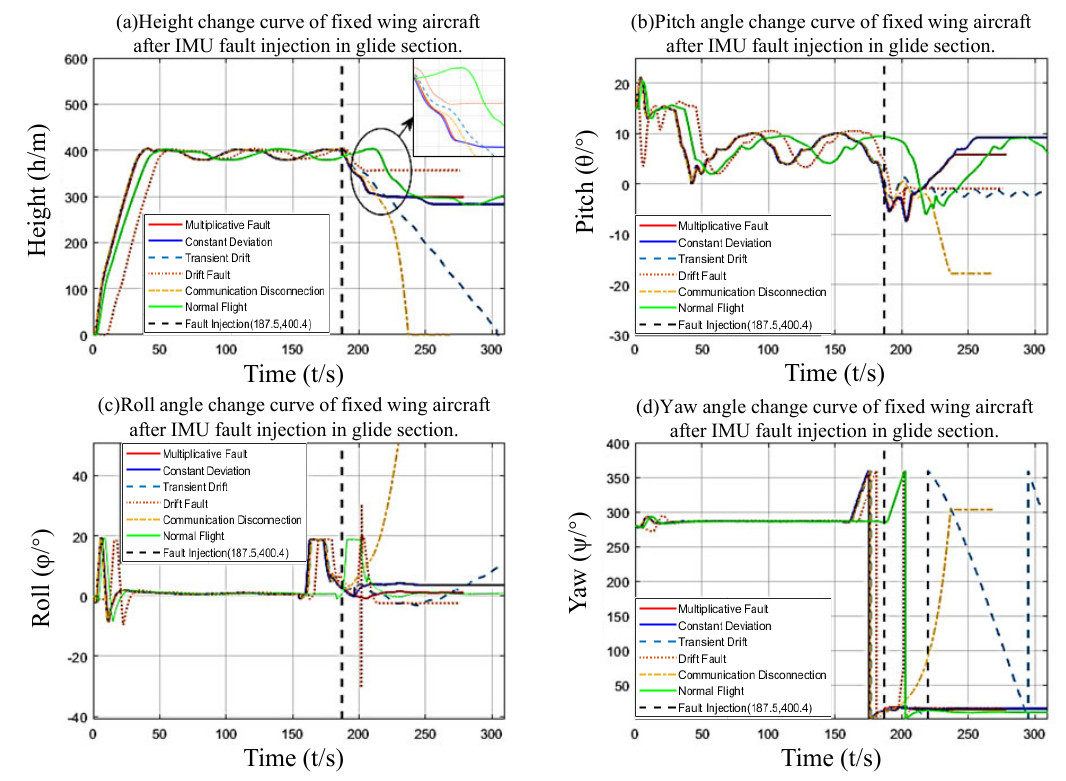}
	\caption{Impact of IMU fault in glide slope on system status.}
	\label{fig_NO.17}
\end{figure}

The IMU drift fault and communication disconnection in the glide slope impact the normal completion of the task of the UAV, but the UAV can still land and attack. Ride fault, constant deviation fault, and transient signal drift will not affect the flight control of UAVs.

From figure \ref{fig_NO.17}, it can be seen that after a given altitude of 400m, the longitude and latitude of the UAV have little change after IMU drift fault is injected into the glide section, and the control strategy is to maintain the previous rudder. At this time, the output value of the airborne vertical gyroscope is randomly added with a constant compared with the output value of the normal value. The pitch angle maintains the previous angle, then glides at a certain descent speed, unable to complete the established task normally.

\subsection{Actuator Failure}
The size of the steering gear fault injected in hardware in the loop simulation is shown in table \ref{injection_gear}.
\begin{table}[!t]
	\renewcommand{\arraystretch}{1.3}
	\caption{Specific parameters of steering gear fault injection}
	\label{injection_gear}
	\centering
	\begin{tabular}{c c}
		\hline
		
		
		Steering Gear Fault Type&Specific Parameters\\
		\hline 
		Constant deviation&All steering gear angles minus 2 °.\\
		Stuck fault&All steering gears are stuck,\\
		&and the steering gear angle is maintained \\
		&at the last normal flight time\\
		Loose fault&All steering gear gains are reduced to 0.9\\
		Damage fault&All steering gear gains are reduced to 0.8\\
		Abnormal jitter fault&Intermittent fault\\
		\hline
	\end{tabular}
\end{table}

\begin{figure}[!t] 
	\centering
	\includegraphics[width=3.25in]{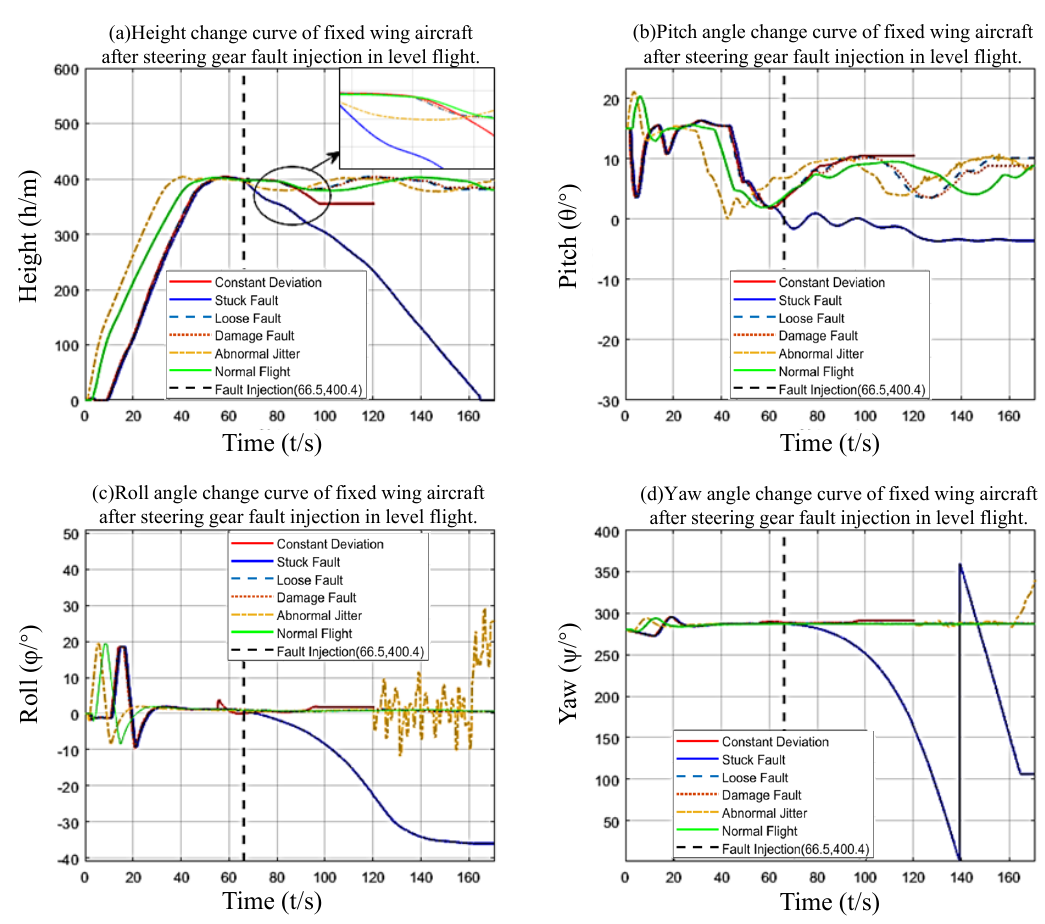}
	\caption{The influence of steering gear fault on system state quantity in level flight phase.}
	\label{fig_NO.18}
\end{figure}

When the UAV runs to the level flight phase, inject a stuck fault in the steering gear channel through the S function as an example, and the system state output is shown in the figure above. When the jamming fault occurs, there is always a control angle for the steering gear, which constantly changes the three-axis attitude angle. The signal difference with the normal flight state increases continuously, leading to the UAV being out of control, causing severe consequences.

The steering gear jamming fault in the level flight phase impacts the normal completion of the mission of the UAV, but the UAV can still land and attack. Constant deviation fault, loose steering gear, damaged steering gear, and abnormal jitter will not affect the flight control of a UAV.

It can be seen from figure \ref{fig_NO.18} that after a given altitude of 400m, the longitude and latitude of the UAV have little change before and after the injection of the steering gear jamming fault in the level flight phase. The three-axis attitude angles are significantly reduced due to the influence of the steering gear jamming. The pitch angle is reduced, the roll angle is reduced, the yaw angle is reduced, the UAV lowers its head, and the altitude begins to decline. The UAV failed to resist the jamming of the steering gear, and the flight control computer could not control the UAV to adjust its attitude. The airspeed of the UAV gradually increased from 40m/s under the effect of gravity and finally hit the ground at a speed of 50m/s, unable to complete the task normally.

\begin{figure}[!t] 
	\centering
	\includegraphics[width=3.25in]{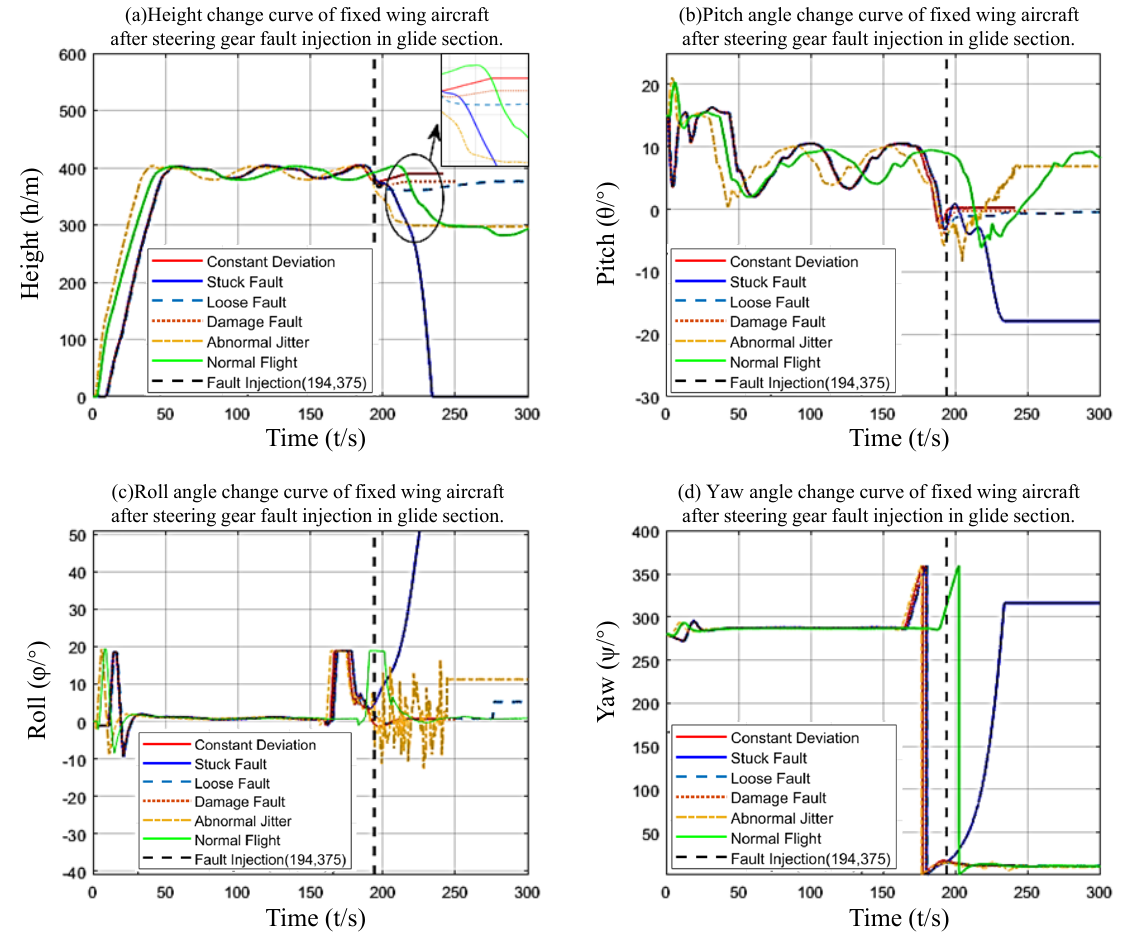}
	\caption{The influence of steering gear failure on system state quantity in glide slope.}
	\label{fig_NO.19}
\end{figure}

The steering gear jamming fault in the glide section impacts the expected completion of the task of the UAV, but the UAV can still land and attack. Constant deviation fault, loose steering gear, damaged steering gear, and abnormal jitter will not affect the flight control of a UAV.

It can be seen from figure \ref{fig_NO.19} that after a given altitude of 400m, the longitude and latitude of the UAV do not change much before and after the injection of the steering gear jamming fault in the glide section, the pitch angle decreases by 10 °, the roll angle decreases, the yaw angle decreases, the UAV lowers its head, and the altitude starts to decline. The UAV failed to resist the interference of the jammed steering gear, and the flight control computer could not control the UAV to adjust its attitude, so it could not complete the task typically.

\section{Damage analysis and failure comparison}
After the fault injection test on the UAV model, relevant test data are obtained, test data and data are sorted out for calculation and analysis, and test results are comprehensively analyzed.

\begin{figure}[!t] 
	\centering
	\includegraphics[width=3.25in]{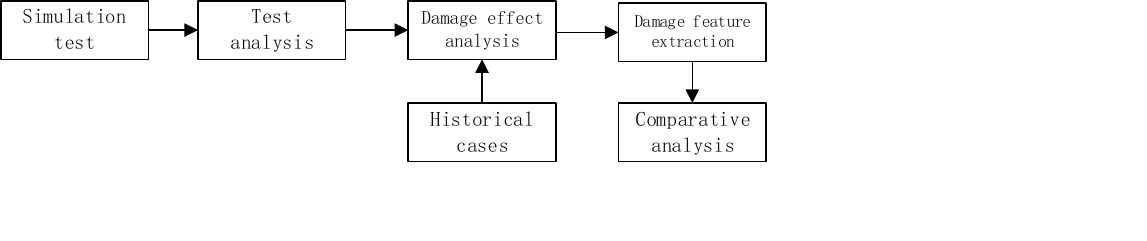}
	\caption{The process of damage effect analysis.}
	\label{fig_NO.20}
\end{figure}

The UAV damage effect analysis process is shown in figure \ref{fig_NO.20}, including:

\begin{enumerate}
	\item Damage effect analysis: analyze the impact of each failure mode on UAV flight and mission/control based on historical cases and test data.
	\item Damage feature extraction: analyze and obtain UAV fault state features.
	\item Comparative analysis: based on the fault state characteristics, analyze and judge the fault location, the impact on system performance, and the development trend of the fault to reverse the fault cause through the fault performance and provide technical support for the accurate strike and effectiveness analysis.
\end{enumerate}

\subsection{GPS Fault Feature Extraction}
GPS deception jamming can locate the receiver in the wrong position, which is harmful. The basic idea of deception jamming is to change the input conditions of the location equation and input the wrong input conditions into the location equation to obtain the wrong solution. Therefore, the characteristic of GPS deception jamming fault is that it cannot maintain the previous flight state at the moment of fault injection, and it turns into an abnormal glide section with a larger glide slope. The roll angle and roll angle rate are rapidly and intensively oscillating. The throttle is reset to zero, which makes the UAV unable to complete the task typically and damages the UAV platform. After receiving the fault, the flight control enters the level flight control strategy, the pitch angle is maintained, the lateral mode remains the same as the navigation state, glides forward, and the altitude is gradually reduced, and finally lands to attack. The GPS deception jamming fault impacts the target of the UAV's surveillance and strike mission, but it can still attack.

\subsection{IMU Fault Feature Extraction}
IMU, namely inertial measurement unit, is used to measure the three-axis attitude angle (or angular rate) and acceleration of objects. IMU provides attitude and angular rate data for UAVs in the flight control system. IMU communication disconnection is harmful, and UAVs cannot obtain attitude and angular rate data. During fault injection, the throttle will return to zero, the altitude will drop, the UAV will bow, roll right to roll left, the yaw angle will fall, the roll angle and yaw angle will drop rapidly, and the roll angle rate and yaw angle rate will drop while shaking. It has an impact on the UAV task and the UAV body platform. IMU drift fault is harmful. When the fault is injected, the throttle will return to zero, the altitude will drop, the UAV will bow, roll right to roll left, the yaw angle will fall, and the roll angle and yaw angle will drop rapidly after vibration. It will affect the UAV task and the UAV body platform.The IMU transient signal drift has little impact on the UAV mission but no impact on the UAV platform. The feature is that after fault injection, the three-axis attitude angle and angular attitude rate suddenly change. The airspeed rises and then drops, and the throttle falls and then rises.

\subsection{Fault Feature Extraction of Steering Gear Faults}
The characteristics of the constant deviation of the steering gear are that it does not affect the airspeed. The three-axis attitude angle changes less, and there is no damage to the UAV platform. The UAV can generally complete the established task, and the aircraft is yaw free. The constant deviation of the injection steering gear in the climb phase does not affect the climb. The injection failure in the level flight phase makes the aircraft maintain level flight at a lower altitude, and the injection failure in the glide phase makes the aircraft turn to level flight.
The characteristic of the steering gear jamming fault is that the steering gear jamming fault is injected in a specific flight phase. The control surface and flight characteristics of the previous flight phase will be maintained in the next flight phase, reflected explicitly in the delay of damage to the UAV platform. They will interfere with the regular progress of the aircraft mission. For example, the UAV with the injection steering gear stuck in the climb phase can generally complete the climb. Still, the steering gear cannot normally rotate after entering the level flight phase, resulting in the UAV lowering its head and altitude.
The looseness of the steering gear is similar to the damage to the steering gear. After receiving the fault, the model cannot fully respond to the steering command sent by the flight control and cannot reach the target deflection angle given by the flight control, which increases the difficulty of flight control. There is no noticeable feature. The pitch angle changes abnormally and fluctuates up and down. The aircraft can normally complete the task with minor UAV platform damage. The three-axis attitude angle changes steadily.
The abnormal jitter fault of the steering gear belongs to the odd departure of the steering gear, characterized by sharp changes in the roll angle, little impact on the altitude trajectory, burrs in the pitch angle change curve, and no obvious airspeed and altitude anomalies. It does not affect the UAV platform, and the aircraft can complete the task usually.

\section{Conclusion}
Aiming at the problem of lack of fault data of UAVs, this paper proposes hardware in the loop simulation method based on the UAV digital simulation model to generate fault data. First, a Simulink simulation model of the “Coyote" UAV is built according to the aerodynamic shape, and structural parameters of the American “Coyote" UAV, and then the dynamic characteristics of the model are verified. The matching degree between the simulation model and the real UAV is verified using actual flight data. Then, based on the simulation model, fault modeling and fault injection of sensors and actuators are completed to generate fault data. All digital simulation tests in this paper are carried out on the MATLAB 2018b platform.

The laboratory independently develops the hardware and software of this hardware in the loop simulation system. It has a broad versatility for the UAV platform. The experimental process based on multiple flights and airfoil tests is convenient. According to the different fault sizes, injection nodes, and fault types, a series of fault data are generated through hardware in the loop simulation, which provides a practical way for the damage effect analysis and feature extraction of the UAV Efficient and reliable solution. The rapid and effective evaluation of UAV status is necessary to improve weapon attack effectiveness. The damage effect evaluation results of UAVs serve as the basis for the commander's decision-making, quickly judge the damage of UAV, provide criteria for transferring the next target or continuing to attack, and achieve accurate fire control and efficient attack.


\section*{Acknowledgment}
This research was supported by Northwestern Polytechnical University and Xi’an Lyncon Electronic Sci. \& Tech. Co., Ltd.



%

\end{document}